# Hybrid Virtual Space Vector Modulation With Extended Power Factor Range for Coupled Eight-Switch Three-Phase Three-Level Inverter

Xiaojun Deng, Hongliang Wang, Tianfu Sun, Yang Jiang, Yibo Si, Wenbo Tang, Mingrui Yang and Xiumei Yue
College of Electrical and Information Engineering, Hunan University, Changsha 410082, Hunan Province, China
wanghl123@hnu.edu.cn

**ABSTRACT:** With the development of photovoltaic (PV) and motor drive, the coupled eight-switch three-phase three-level inverter (TP-TLI) is receiving more attentions. However, the previous modulation strategies for coupled eight-switch TP-TLI are suitable for high power factor (PF), which cannot satisfy the requirement for PV and motor drive. Aiming at this, this paper proposes a hybrid virtual space vector modulation (HVSVM) for the coupled eight-switch TP-TLI with extended PF range. The mechanism of PF limitation is analyzed in deep. Two virtual vectors are constructed in each sector, and three basic VSVMs with different PF range are proposed according to the number of virtual small vectors used. Combined with three VSVMs, the HVSVM is proposed by detecting the three-phase current polarities. The PF angle range of the HVSVM is extended from about [-30°, 30°] to [-90°, 90°]. Besides, an active neutral-point voltage control (NPVC) is also presented by using two coefficients to adjust the dwell times of the two pairs of redundant small vectors. This approach considers the NP current and the derivation of NPV, and thereby it is accurate to control the NPV. The correctness and feasibility of the proposed HVSVM and active NPVC are verified by experiments through a 3 kVA prototype.

**KEY WORDS:** Inverter, power factor (PF), neutral-point voltage (NPV), virtual space vector modulation (VSVM).

## I. INTRODUCTION

Due to the improved power capacity and total harmonic distortion (THD), three-phase three-level inverters (TP-TLIs) have been key equipment in photovoltaic (PV) application, motor drives, and other applications [1]-[4]. Among all the TP-TLIs, the coupled ten- and eight-switch TP-TLI are two of the most attractive topologies as their switch counts are lower than other similar topologies [5]-[7]. The circuit diagrams of them are shown in Fig. 1. The current stresses on all the switches for them are the same as those for traditional TP-TLIs, and the voltage stresses on switches for them are the same as those for T-type TP-TLI [8]. However, the switch counts are reduced to ten and eight for the coupled ten- and eight-switch TP-TLI, respectively. Owing to these, these two topologies have been applied in PV and motor drives.

However, the eight-switch TP-TLI still faces two serious issues including limited power factor range and unbalanced neutral-point voltage. For the first issue, seen from Fig. 1(b), diodes $D_1$ and $D_2$ and switches $S_1$ and $S_2$ consist a coupled unit, which are shared for three-phase.

This work was supported by the National Natural Science Foundation of China Under Grant 51977069, the First Key Research and JieBang Headed Program, Hunan Province, China Under Grant 2021GK1250, the National Natural Science Foundation Youth Project of China Under Grant 52107195, and the Science and Technology Planning Project of Suzhou City, China Under Grant SZS2022015.

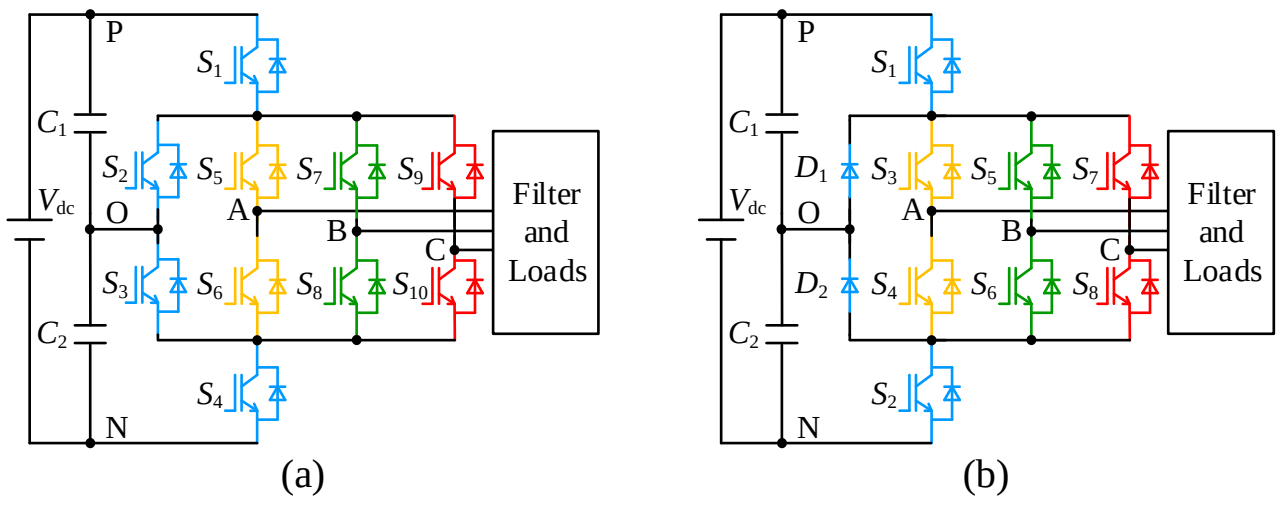

Fig. 1. (a) Coupled ten-switch TP-TLI. (b) Coupled eight-switch TP-TLI.

Hence, the polarity of phase current flowing though $D_1$ and $D_2$ will be limited. In [6], an improved sinusoidal pulse-width modulation (SPWM) was presented for this topology. However, it is viable for PF $\in$ [0.66, 1]. Besides, the PF should be greater than or equal to 0.84 if the zero sequence voltage (ZSV) is injected to achieve the same dc voltage utilization as space vector modulation (SVM). However, the above methods cannot satisfy the PF range for PV and motor drives, and thereby PF range extension is necessary for the coupled eight-switch TP-TLI.

For another issue, the unbalanced NPV increases the voltage stresses of dc capacitors and switches, and even damages them [6]-[7]. To address this issue, numerous efforts have been paid in ten- and eight-switch TP-TLI [6]-[12]. In [6], the NPVC based on SPWM is realized by injecting ZSV into the reference. In [7], voltage balancing band is defined, and a coefficient with three fixed values is used to adjust the dwell time of redundant small vector based on SVM when the NPV is out of this band. A novel SVM, which adjusts the dwell times of non-redundant small vector was proposed for controlling the NPV in [9]. However, there are low frequency ripples in NPV for the above methods based on SPWM and SVM. Consequently, virtual SVM (VSVM), which uses line combination of basic vectors to provide zero NP current, is an attractive candidate [19]. Two novel VSVMs with optimized NPVC capability were proposed for the ten-switch TP-TLI in [11] and [12]. When they are applied for the eight-switch TP-TLI, the PF should be greater than or equal to 0.867. Hence, it is still essential to simultaneously control the NPV and extend the PF range for the eight-switch TP-TLI.

To achieve these targets, this paper proposed a hybrid VSVM (HVSVM) and an active NPVC method for the eight-switch TP-TLI. It contains three basic VSVMs, whose union of PF range could cover the PF range for inverting operation. The proposed HVSVM can adaptively transits from these three basic VSVMs based on the three-phase current polarities. Consequently, the PF of the proposed HVSVM could be extended to [0, 1]. Besides, the active NPVC considers the NP current produced by each virtual vector, and thereby it is accurate to control the NPV under all modulation indexes and PF. This organization of this paper is as follows. Section II presents the operation principles of the eight-switch TP-TLI. Section III describes the proposed HVSVM and its active NPVC method. Section IV gives the experimental results. Finally, the conclusion is shown in Section V.

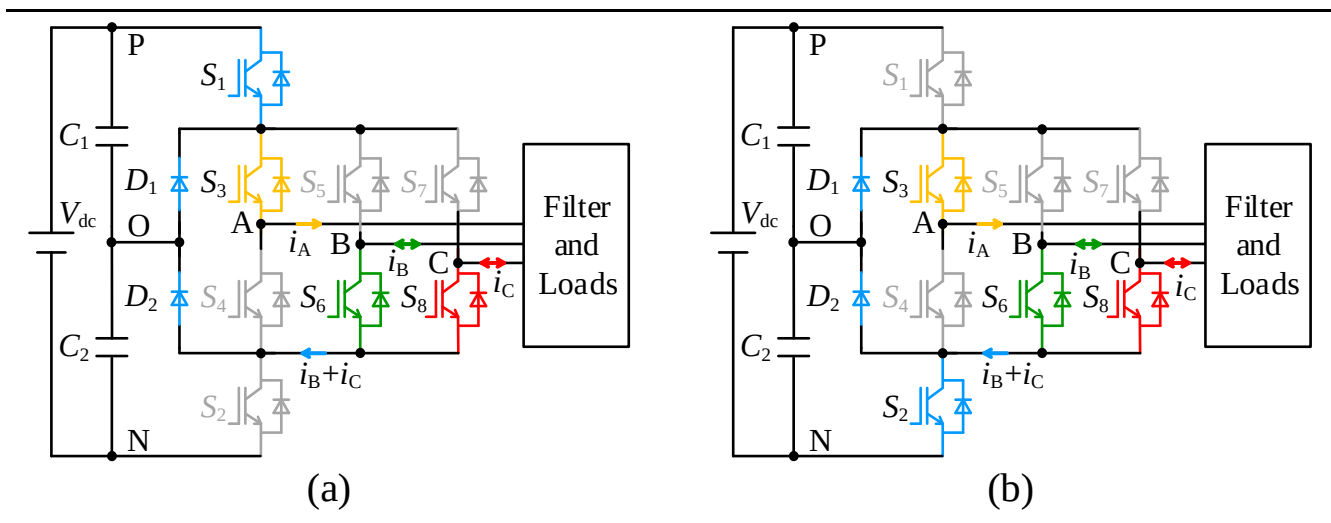

Fig. 2. Circuit examples for the eight-switch TP-TLI. (a) 211. (b) 100.

TABLE I SWITCHING STATES OF THE EIGHT-SWITCH TP-TLI

| Space vectors | Switches combinations | Magnitudes | Classification |
|---|---|---|---|
| 000 ($V_{0L}$) | {2, 4, 6, 8} | 0 | Zero vector |
| 111a to 111h ($V_{0M}$) | {1, 4, 6, 8}, {4, 6, 7}, {4, 5, 8}, {3, 6, 8}, {4, 5, 7}, {3, 6, 7}, {3, 5, 8}, {2, 3, 5, 7} | | |
| 222 ($V_{0U}$) | {1, 3, 5, 7} | | |
| 100 ($V_{1L}$), 211 ($V_{1U}$) | {2, 3, 6, 8}, {1, 3, 6, 8} | $V_{dc}/3$ | Small vectors |
| 110 ($V_{2L}$), 221($V_{2U}$) | {2, 3, 5, 8}, {1, 3, 5, 8} | | |
| 010 ($V_{3L}$), 121 ($V_{3U}$) | {2, 4, 5, 8}, {1, 4, 5, 8} | | |
| 011 ($V_{4L}$), 122 ($V_{4U}$) | {2, 4, 5, 7}, {1, 4, 5, 7} | | |
| 001 ($V_{5L}$), 112 ($V_{5U}$) | {2, 4, 6, 7}, {1, 4, 6, 7} | | |
| 101 ($V_{6L}$), 212 ($V_{6U}$) | {2, 3, 6, 7}, {1, 3, 6, 7} | | |
| 200 ($V_7$), 220 ($V_8$) | {1, 2, 3, 6, 8}, {1, 2, 3, 5, 8} | $2V_{dc}/3$ | Large vectors |
| 020 ($V_9$), 022 ($V_{10}$) | {1, 2, 4, 5, 8}, {1, 2, 4, 5, 7} | | |
| 002 ($V_{11}$), 202 ($V_{12}$) | {1, 2, 4, 6, 7}, {1, 2, 3, 6, 7} | | |

## II. OPERATION PRINCIPLES OF THE COUPLED EIGHT-SWITCH TP-TLI

### A. Topology and Its Operation Principles

Fig. 1(b) depicts the circuit configuration of the eight-switch TP-TLI. As seen, it contains eight switches ($S_1$-$S_8$) and two diodes ($D_1$-$D_2$). The overall dc-link voltage is divided by two capacitors $C_1$ and $C_2$. The reference is set at the negative dc-link N. Hence, three voltage levels $V_{dc}$, $V_{dc}/2$, and 0 can be obtained, and are respectively denoted as 2, 1, and 0. The positive direction (denoted as +) of phase current is defined as the flow from dc side to loads.

All the switching state for the coupled eight-switch TP-TLI are shown in Table I. The outputs of each switching state are denoted as “ABC”. For example, switching state 211 indicates that phases A, B, and C respectively output levels 2, 1, and 0. It is achieved by switching $S_1$, $S_3$, $S_6$, and $S_8$ into on state, as shown in Fig. 2(a), in which the corresponding on switches and off switches are marked in color lines and gray lines, respectively. Taking switching state 100 shown in Fig. 2(b) as another example, phases A, B, and C output levels 1, 0, and 0, respectively. It can be achieved if switches $S_2$, $S_3$, $S_6$, and $S_8$ are all turned on. 111a to 111h are record as one switching state since they have the same outputs.

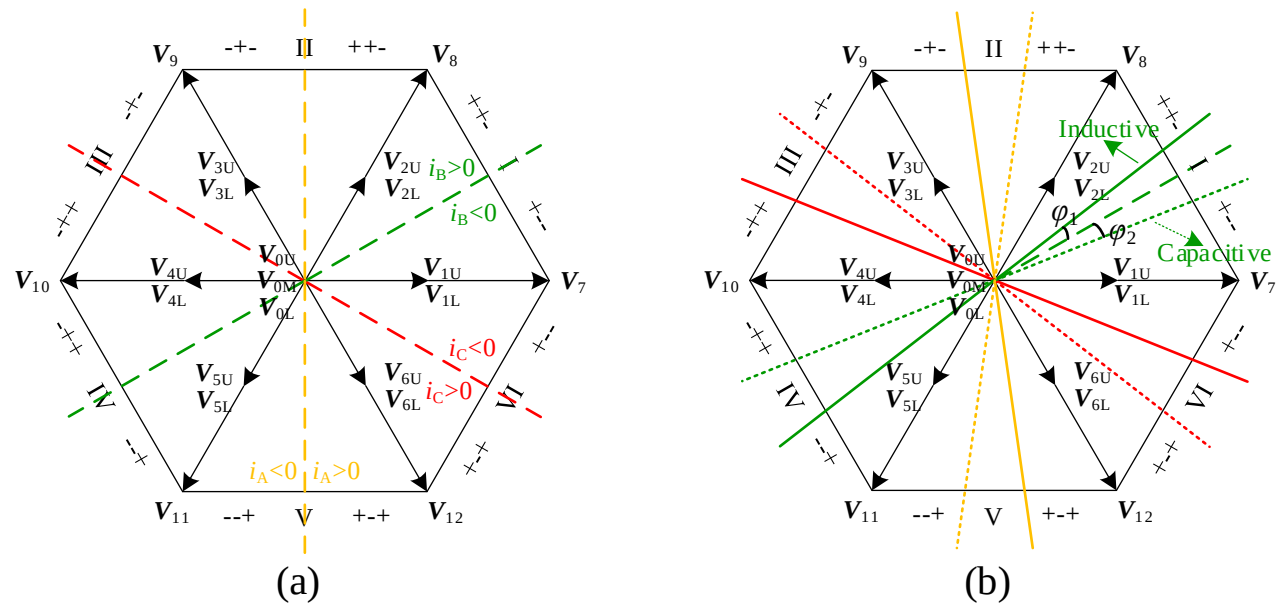

Fig. 3. Space vector diagram and current directions of the eight-switch TP-TLI. (a) Under UPF. (b) Under lagging and leading PF.

### B. PF Limitation Analysis

Fig. 3(a) shows the space vector diagram of the eight-switch TP-TLI. As can be observed, there are 21 space vector including six large vectors, 12 small vectors, and 3 zero vectors. Besides, some vectors have the same spatial position, and they are called as redundant vectors. For better illustrating the PF limitation of this topology, Fig. 3(a) also depicted the three-phase current polarities at $\varphi$ = 0°, where $\varphi$ is the PF angle. These polarities are rotated clockwise a PF angle with respect to the coordinate origin for $\varphi > 0$, whereas they are rotated anticlockwise for $\varphi < 0$.

It is seen from Fig. 1(b) that diodes $D_1$ or $D_2$ are shared for three-phase. Owing to the unipolar conductivity of diode, the polarity of phase current flowing through them are limited. Taking 100 as an example, only phase current $i_A$ flows through $D_1$. This implies the polarity of $i_A$ is positive (+). Considering three-phase symmetry, the polarities for three-phase currents are +-+, +--, ++-, as recorded in Table II. Similar analysis can be applied for the other switching states. Table II lists all the switching states with limited current polarities. It is seen that the polarities of all the small vectors and most zero vectors are limited.

Based on these, the PF range of the eight-switch TP-TLI can be analyzed. For example, 211, 100, 221, 110 are all used in [8]-[12], and thereby the current limitations are +-- and ++-. They will be appeared for PF ∈ [0.867, 1], and thereby the PF range is [0.867, 1] for [8]-[12].

## III. THE PROPOSED HYBRID VSVM FOR THE EIGHT SWITCH TP-TLI

### A. Virtual Vector Construction

Two virtual vectors are constructed in each sector to expansion the PF range and balance the NPV of the eight-switch TP-TLI. For example, the two virtual vectors in sector I is shown in (1), where coefficient $k_1$ or $k_2$ are used to control the NPV by regulating the dwell times of each pairs of redundant vectors. The NP currents generated by these two virtual vectors are given in (2). The NP current can be positive, 0, and negative since $k_1$ and $k_2$ ∈ [-1, 1].

$$\begin{cases} \boldsymbol{V}_{1.1} = ((1-k_1)\boldsymbol{V}_{1L} + (1+k_1)\boldsymbol{V}_{1U})/2 \\ \boldsymbol{V}_{2.1} = ((1-k_2)\boldsymbol{V}_{2L} + (1+k_2)\boldsymbol{V}_{2U})/2 \end{cases} \quad (1)$$

$$i_{O1.1} = -k_1 i_A,\ \ i_{O2.1} = k_2 i_C \quad (2)$$

The current limitations of each virtual vector and its components are the same, as seen in Table II and Table III. Fig. 4(a) gives the virtual space vector diagram. As seen, there are one zero vector, six virtual small vectors, and six large vectors in Fig. 4(a).

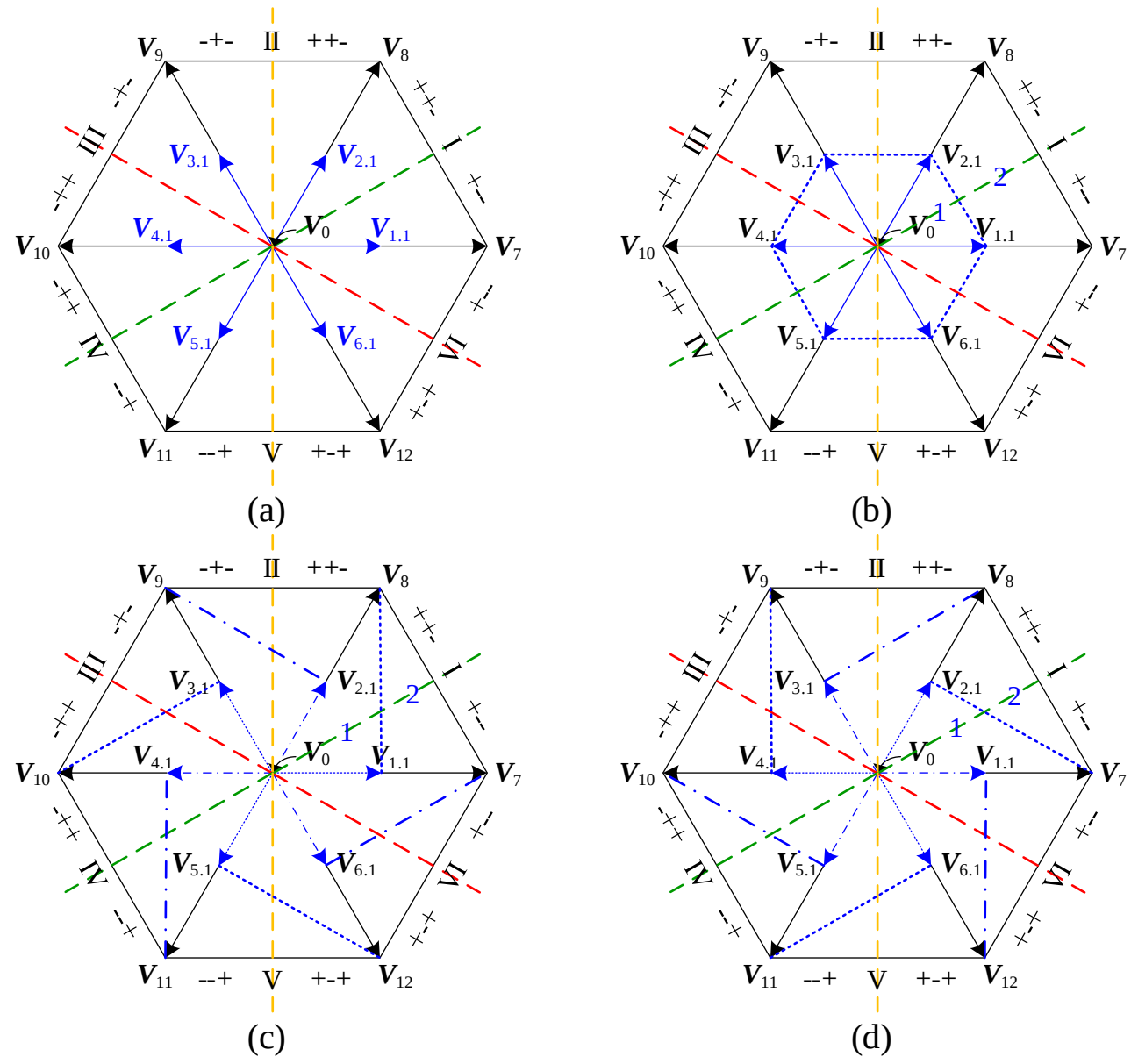

Fig. 4. Virtual space vector diagram and modualtion strategy. (a) Virtual space vector diagram. (b) VSVM A. (c) VSVM B. (d) VSVM C.

TABLE II CURRENTS RESTRICTIONS ON SWITCHING STATES

| Switching state | Currents directions | Switching state | Currents directions |
|---|---|---|---|
| 111d/100/211 | +-+, +--, ++- | 111g/110/221 | +--, ++-, -+- |
| 111c/010/121 | ++-, -+-, -++ | 111e/011/122 | -+-, -++, --+ |
| 111b/001/112 | -++, --+, +-+ | 111f/101/212 | --+, +-+, +-- |

## B. Basic VSVM

According to the utilized virtual small vector in each sector, three basic VSVMs names as VSVM A to VSVM C can be obtained. VSVMA utilizes two virtual small vectors in each sectors. While both VSVM B and VSVM C uses one virtual small vector, but their spatial positions are different. Sector I is taken as an example.

(1) VSVM A: As shown in Fig. 4(b), it uses all the two virtual small vectors, and thereby the three-phase current polarities should satisfy with the intersection of current restrictions of them. Seen from Table III, the intersections of current restrictions are +-- and ++-. They appear for $\varphi \in$ [-30°, 30°]. Hence, the PF of VSVM A ∈ [-30°, 30°].

(2) VSVM B: It uses only one virtual small vector $\boldsymbol{V}_{1.1}$, as seen from Fig. 4(c), and thereby the three-phase current polarities should be +-+, +--, and ++-. They appear for $\varphi \in$ [-90°, 30°]. Hence, the PF of VSVM B ∈ [-90°, 30°].

(3) VSVM C: It uses another virtual small vector $\boldsymbol{V}_{2.1}$, as can be seen from Fig. 4(d), and thereby the three-phase current polarities are +-+, +--, and ++-. They appear for $\varphi \in$ [-30°, 90°]. Hence, the PF of VSVM C ∈ [-30°, 90°].

## C. HVSVM

Here, sector I is taken as an example for better illustrating the proposed HVSVM. The proposed HVSVM divides $\varphi \in$ [-90°, 90°] for inverting operation into three different areas based on the three-phase current polarities.

(1) $\varphi \in$ [-30°, 30°]: In this condition, the possible three-phase current polarities are +-- and ++-. VSVM A to VSVM C are all satisfied with these current polarities, as seen in Fig. 5(a). However, VSVM A provides better output performance in low modulation index. Hence, VSVM A is adopted for $\varphi \in$ [-30°, 30°], as can be seen in Fig. 5(b).

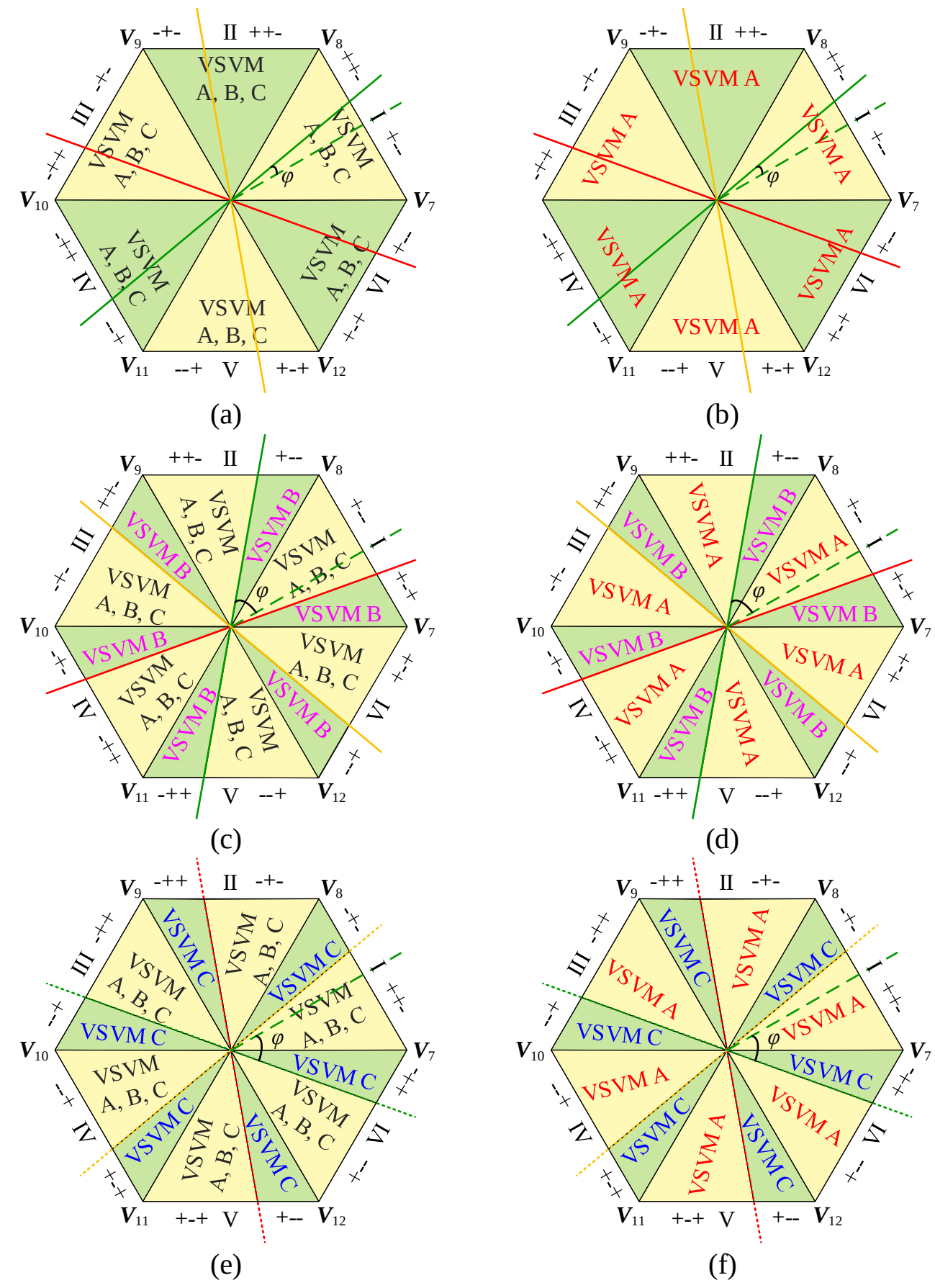

Fig. 5. Illustration of the proposed HVSVM. (a), (b) $\varphi \in$ [-30°, 30°]. (c), (d) $\varphi \in$ [-90°, -30°]. (e), (f) $\varphi \in$ [30°, 90°]. (a), (c), (e) Available VSVM for different PF angle $\varphi$. (b), (d), (f) Adopted VSVM for the proposed HVSVM.

TABLE III CURRENTS RESTRICTIONS FOR VIRTUAL SMALL VECTORS

| Virtual small vectors | Currents directions | Virtual small vectors | Currents directions |
|---|---|---|---|
| $\boldsymbol{V}_{1.1}$ | +-+, +--, ++- | $\boldsymbol{V}_{2.1}$ | +--, ++-, -+- |
| $\boldsymbol{V}_{3.1}$ | ++-, -+-, -++ | $\boldsymbol{V}_{4.1}$ | -+-, -++, --+ |
| $\boldsymbol{V}_{5.1}$ | -++, --+, +-+ | $\boldsymbol{V}_{6.1}$ | --+, +-+, +-- |

(2) $\varphi \in$ [-90°, -30°]: In this area, the possible three-phase current polarities are +-+ and +--. For +-+, only VSVM B can be applied, as seen in Fig. 5(c). While for +--, VSVM A to VSVM C are all satisfied with it, and VSVM A is adopted for its better output performance in low modulation index, as seen in Fig. 5(d).

(3) $\varphi \in$ [30°, 90°]: In this condition, the possible three-phase current polarities are -+- and ++-. For -+-, only VSVM C can be applied, as seen in Fig. 5(e). While for ++-, VSVM A to VSVM C are all satisfied with it, and VSVM A is the preferred solution, as seen in Fig. 5(f).

## D. Active NPVC

The proposed active NPVC can be realized by adjusting the dwell times of the two redundant small vectors of each virtual vector. Here, sector I is taken as an example. In the HVSVM, if VSVM A is detected, the dwell times for two pairs of redundant small vectors $\boldsymbol{V}_{1L}$ and $\boldsymbol{V}_{1U}$, and $\boldsymbol{V}_{2L}$ and $\boldsymbol{V}_{2U}$ are respectively adjusted by two coefficients $k_1$ and $k_2$, as shown in (1). In this condition, the average NP current $i_{Oav}$ produced by them should be equal to the required NP current $i_{Oref}$, as shown in (3). The required NP current $i_{Oref}$ can be achieved by utilizing the NPV model introduced in

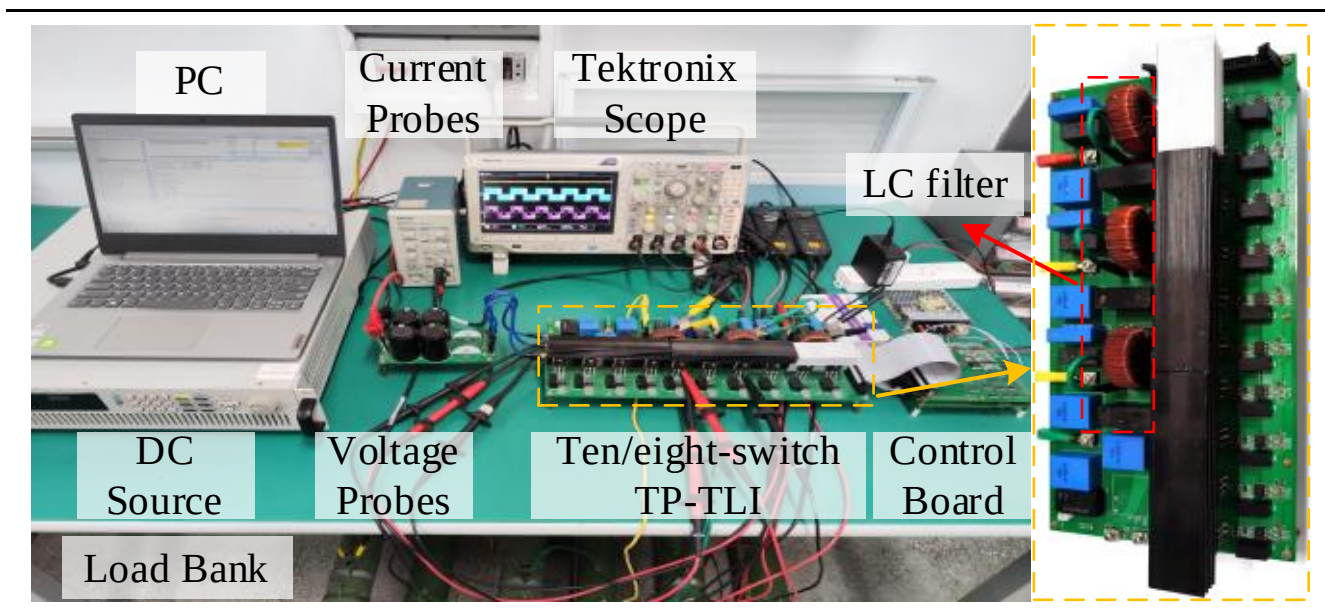

Fig. 6. Experimental prototype for the coupled eight-switch TP-TLI.

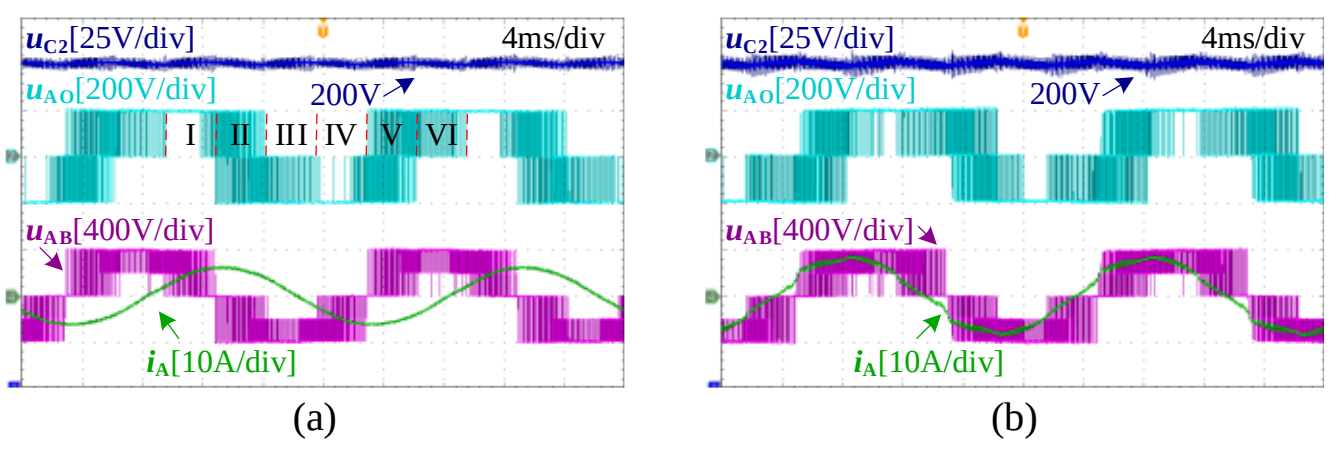

Fig. 7. Steady-state experimental results for the SVM [9]. (a) $m$ = 0.9, $\varphi$ = -70°. (b) $m$ = 0.9, $\varphi$ = 60°.

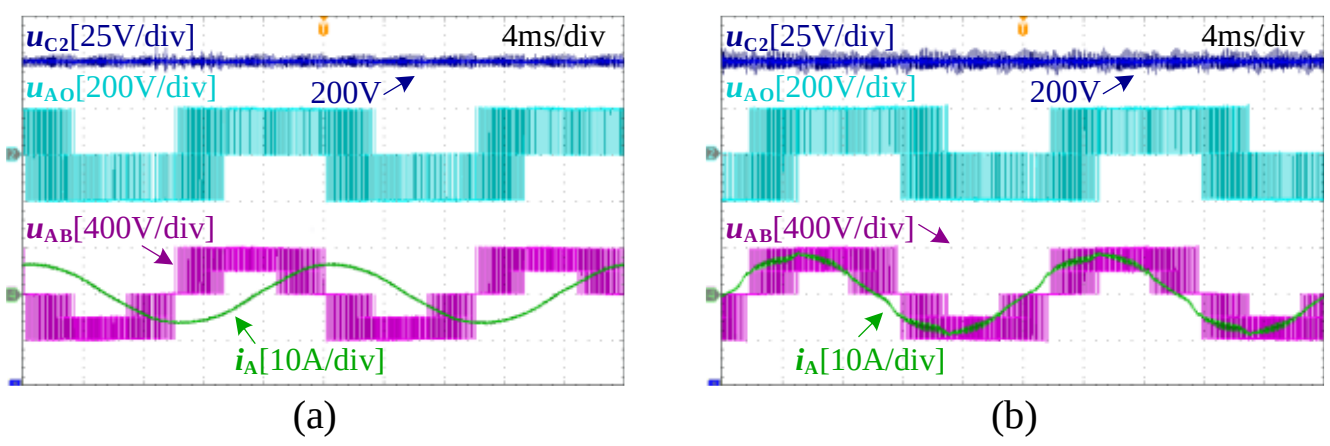

Fig. 8. Steady-state experimental results for the VSVM [10]. (a) $m$ = 0.9, $\varphi$ = -70°. (b) $m$ = 0.9, $\varphi$ = 60°.

TABLE IV EXPERIMENTAL PARAMETERS

| Parameters | Symbols | Values |
|---|---|---|
| input voltage | $V_{dc}$ | 400 V |
| dc-link capacitors | $C_1$ and $C_2$ | 470 uF |
| filter inductor | $L_f$ | 6 mH |
| filter capacitor | $C_f$ | 4.7 uF |
| output frequency | $f_o$ | 50 Hz |
| resistive-inductive load 1 | $RL_1$ | 45 Ω, 70 mH ($\varphi$ = -25°) |
| resistive-inductive load 2 | $RL_2$ | 8 Ω, 70 mH ($\varphi$ = -70°) |
| resistive-capacitive load 1 | $RC_1$ | 45 Ω, 140 uF ($\varphi$ = 25°) |
| resistive-capacitive load 2 | $RC_2$ | 13 Ω, 140 uF ($\varphi$ = 60°) |
| average switching frequency | $f_{avs}$ | 4 kHz |
| devices | $S_1$–$S_2$ | IKW40N60 |
| | $S_3$–$S_8$ | IKW40N120 |
| | $D_1$–$D_2$ | anti-parallel diode of IKW40N60 |

[9], [11], and [12].

$$i_{\text{Oref}} = i_{\text{Oav}} = (k_1 t_{1.1}(-i_{\text{A}}) + k_2 t_{2.1} i_{\text{C}}) / T_{\text{s}} \qquad (3)$$

Assuming the absolute value of $k_1$ equals that of $k_2$, these two coefficients can be achieved as (4) with the considering of the polarity of $-i_A$ and $i_C$ in (3).

$$\begin{cases} k_2 = k_1 = i_{\text{Oref}} T_{\text{s}} / (-t_{1.1} i_{\text{A}} + t_{2.1} i_{\text{C}}), -i_{\text{A}} i_{\text{C}} > 0 \\ -k_2 = k_1 = i_{\text{Oref}} T_{\text{s}} / (-t_{1.1} i_{\text{A}} - t_{2.1} i_{\text{C}}), -i_{\text{A}} i_{\text{C}} \le 0 \end{cases} \qquad (4)$$

$k_1$ and $k_2$ can be similarly solved in other regions. Since the active NPVC is based on VSVM and it considers the NP current produced by each virtual vector, it is accurate to control the NPV under all modulation indexes and PF.

## IV. EXPERIMENTAL VERIFICATIONS

A 3 kVA experimental prototype shown in Fig. 6 is built to verify the HVSVM and active NPVC. The detail experimental parameters are given in Table IV. Three groups of experiments are conducted. They are steady-state operation, NPV recovery operation, and transient operation.

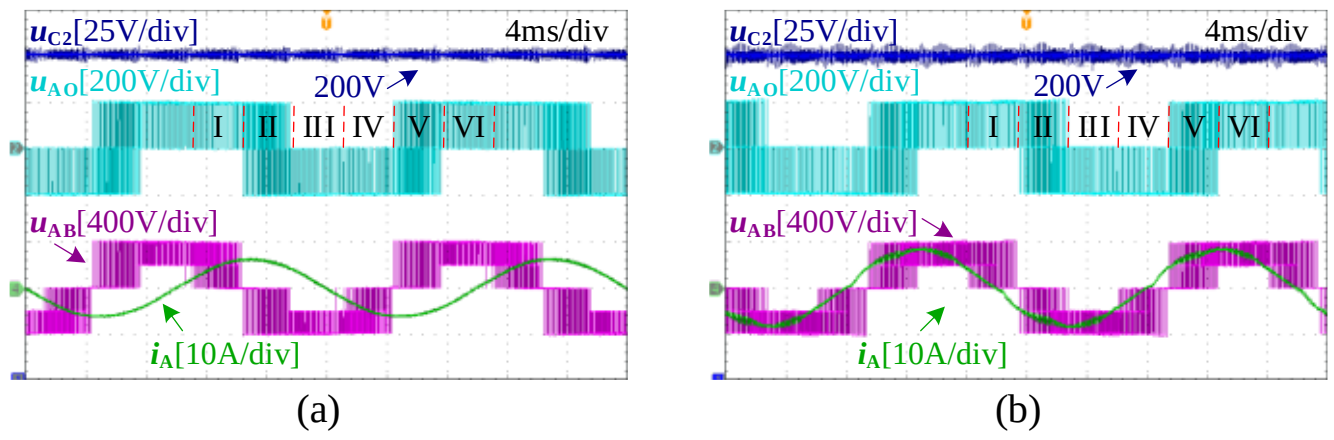

Fig. 9. Steady-state experimental results for the proposed HVSVM. (a) $m$ = 0.9, $\varphi$ = -70°. (b) $m$ = 0.9, $\varphi$ = 60°.

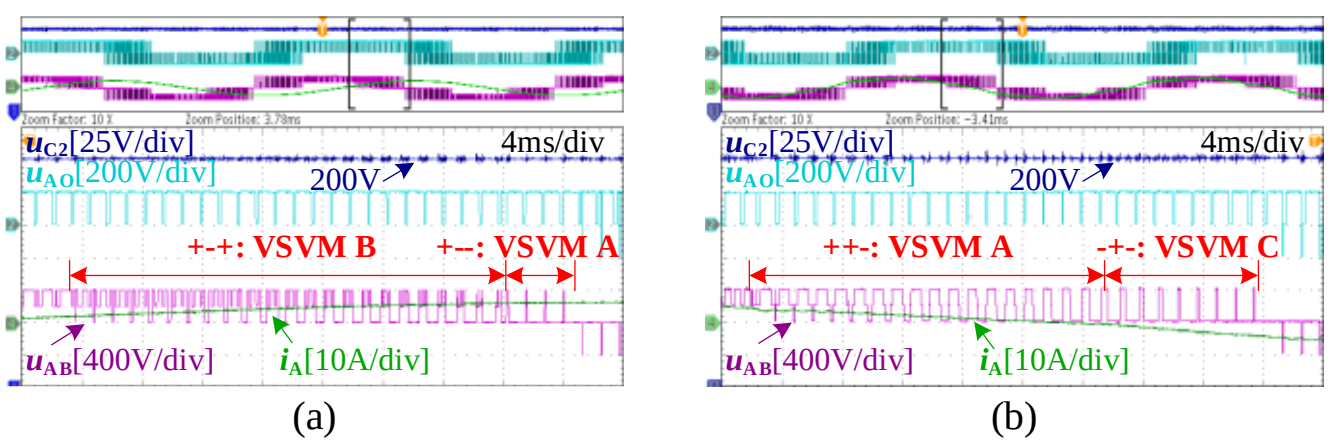

Fig. 10. Partial detail enlargement in sector I. (a) Fig. 9(a). (b) Fig. 9(b).

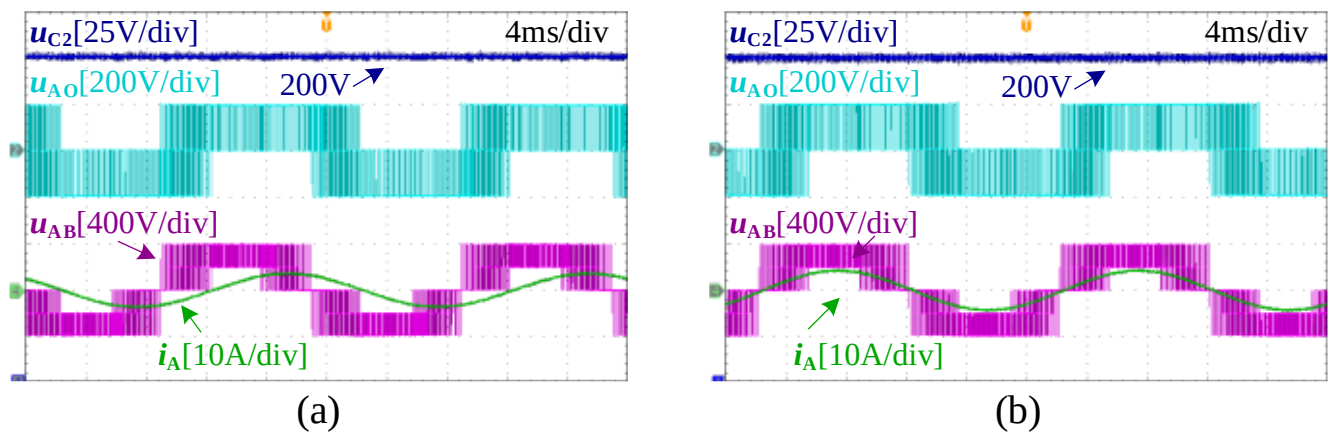

Fig. 11. Steady-state experimental results for the proposed HVSVM. (a) $m$ = 0.9, $\varphi$ = -25°. (b) $m$ = 0.9, $\varphi$ = 25°.

TABLE V RMS, AND HARMONIC CONTENTS COMPARISONS OF $M$ = 0.9 FOR DIFFERENT METHODS

| $\varphi$ | Strategies | $u_{AB}$ RMS (V) | $u_{AB}$ THD u(%) | $u_{AB}$ 5th | $u_{AB}$ 7th | $i_A$ THD i(%) | $i_A$ 5th | $i_A$ 7th |
|---|---|---|---|---|---|---|---|---|
| -25° | HVSVM | 254.3 | 53.9 | 0.95 | 0.75 | 2.17 | 0.67 | 0.26 |
| | SVM [9] | 255.0 | 54.4 | 0.69 | 0.47 | 2.16 | 0.85 | 0.31 |
| | VSVM [10] | 253.4 | 54.0 | 0.70 | 0.50 | 2.07 | 0.66 | 0.28 |
| 25° | HVSVM | 253.7 | 53.5 | 0.15 | 0.23 | 2.42 | 0.36 | 0.19 |
| | SVM [9] | 255.2 | 53.9 | 0.16 | 0.44 | 2.42 | 0.58 | 0.74 |
| | VSVM [10] | 253.1 | 54.0 | 0.79 | 0.20 | 2.46 | 0.45 | 0.21 |
| -70° | HVSVM | 254.6 | 53.0 | 0.56 | 0.74 | 2.11 | 0.87 | 0.50 |
| | SVM [9] | 262.3 | 53.9 | 1.66 | 1.20 | 2.26 | 1.21 | 0.57 |
| | VSVM [10] | 263.2 | 52.0 | 1.65 | 0.88 | 2.43 | 1.23 | 0.63 |
| 60° | HVSVM | 254.3 | 52.9 | 0.39 | 0.31 | 6.47 | 2.80 | 1.72 |
| | SVM [9] | 258.9 | 55.1 | 2.34 | 1.82 | 9.17 | 5.45 | 4.78 |
| | VSVM [10] | 256.2 | 54.0 | 1.83 | 0.85 | 8.49 | 5.16 | 2.45 |

### *A. Steady-State Operation*

Fig. 7 to Fig. 9 show the steady-state results for the SVM [9], VSVM [10], and the proposed HVSVM under low PF at $m$ = 0.9, where $u_A$, $u_{C2}$, $u_{AB}$, and $i_A$ respectively denote the phase voltage, NPV, line voltage, and phase current. It is seen that the NPV of these three methods can be controlled around its nominal voltage 200 V. There are low frequency ac ripples in SVM [9], whereas there are no ac ripples for the VSVM [10] and the proposed HVSVM. The above experimental curves imply the feasibility and effectiveness of the proposed active NPVC.

For the PF operation range, it is seen that the currents for SVM [9] and VSVM [10] are all distorted under low PF, whereas those for the proposed HVSVM are sinusoidal. In addition, Fig. 10(a) and (b) give the detail enlargement of sector I for Fig. 9(a) and (b), respectively. Seen in Fig. 10(a), when the polarities of three-phase currents are changed from +-+ to +--, the adopted VSVM is changed from VSVM B to VSVM A. For Fig. 10(b), the adopted

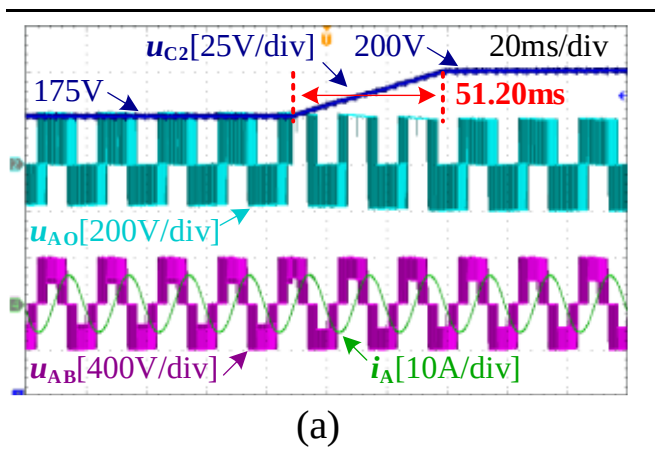

(a)

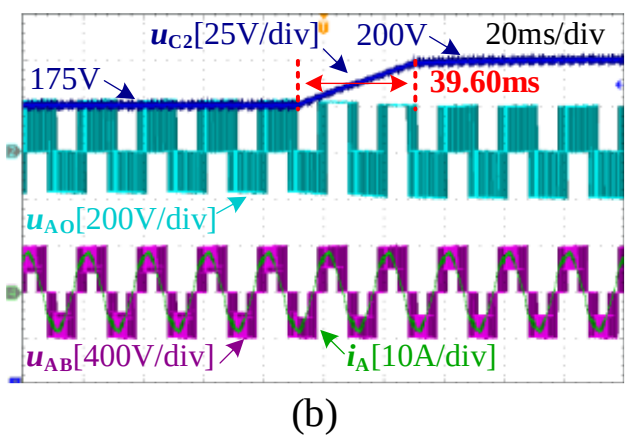

(b)

Fig. 12. NPV recovery process with 50 V offset between two dc capacitors voltages for the proposed VSVM. (a) $m$ = 0.9, $\varphi$ = -70°. (b) $m$ = 0.9, $\varphi$ = 60°.

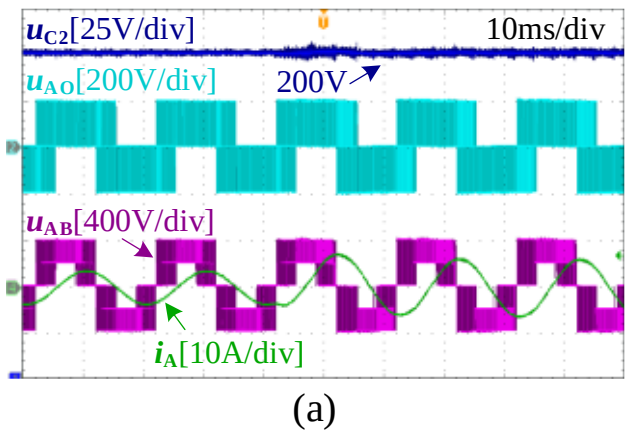

(a)

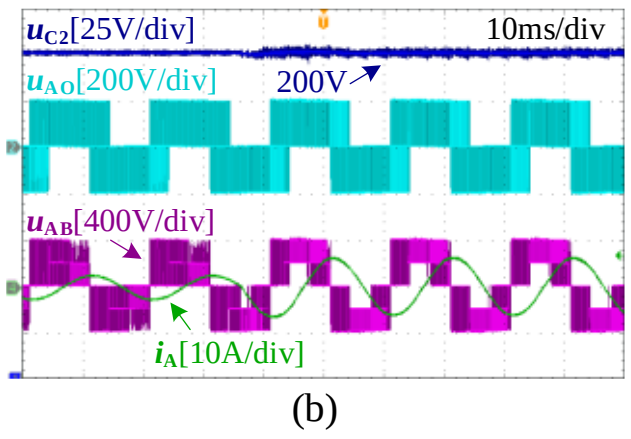

(b)

Fig. 13. Dynamic response for the propsoed VSVM. (a) $\varphi$ is changed from -25° to -70°, $m$ = 0.9. (b) $m$ is changed from 0.4 to 0.9, $\varphi$ = -70°.

VSVM is changed from VSVM A to VSVM C as the current polarities are changed from ++- to -+-. All these indicate the correctness of the proposed HVSVM. Fig. 11 gives the results of the proposed HVSVM under high PF. As seen, the currents are still sinusoidal in shape. Besides, the NPV can be also balanced without obvious dc offset and ac ripples. As a result, the proposed HVSVM is suitable for both high PF and low PF.

Table V gives a comparison of these three methods in terms of the root mean square (RMS) value of line voltage, THD and harmonics contents of line voltage $u_{AB}$ and phase current $i_A$, where THDu and THDi are the THD for $u_{AB}$ and $i_A$, respectively. At $m$ = 0.9, the theoretical RMS value for $u_{AB}$ is about 254.6 V. The maximum error of $u_{AB}$ is about 0.4% for the proposed HVSVM, whereas they are about 3.0% and 3.4% for SVM [9] and VSVM [10], respectively. Hence, the proposed HVSVM could reduce the error of line voltage compared with other two methods. Besides, the low order harmonic contents for the proposed HVSVM are also lower than those for SVM [9] and VSVM [10]. Hence, the above steady-state comparison shows that the proposed HVSVM is viable for both high PF and low PF.

### *B. NPV Recovery Operation*

The NPV recovery operation is given in Fig. 12 for different $\varphi$ at $m$ = 0.9. Initially, 50 V offset is added to the voltages of the two dc capacitors. As seen, the NPV is recovered rapidly. In the recovery process, there is no obvious overshoot. The recovery times are 51.20 ms and 39.60 ms for Fig. 12(a) and (b), respectively. Besides, there are no obvious dc offset and ac ripples when the NPV is recovered. It indicates that the proposed active NPVC have well control ability on NPV under different $\varphi$.

### *C. Transient Operation*

To further shows the transient behavior of the HVSVM, Fig. 13 demonstrates the results involving sudden load alterations and swift $m$ modifications. Fig. 13(a) presents the results when $\varphi$ is changed from -25° to -70° at $m$ = 0.9. The line voltages are accordingly changed, and thereby phase current could keep sinusoidal in shape. This is because the proposed HVSVM could detected the three-phase current polarities. Besides, the NPV can be balanced under load alterations. Fig. 13(b) depicts the results when $m$ is changed from 0.4 to 0.9 at $\varphi$ = -70°. In this condition, the line voltages are changed from three levels to five levels, and the phase current escalates in direct proportion to the rise in $m$. Besides, there is no obvious dc offset and ac ripples in NPV. Therefore, the proposed HVSVM shows well dynamic response under transient operation.

## V. CONCLUSION

A HVSVM with extended PF range is proposed for the coupled eight-switch TP-TLI in this paper. The proposed HVSVM are consisted of three basic VSVMs. Each basic VSVM has different PF range. However, their PF union could cover the overall range of inverting operation. Therefore, the proposed HVSVM adaptively switches among these three basic VSVMs based on the three-phase current polarities and the output performance of each VSVM. Compared with other methods, the PF angle range of the proposed HVSVM is extended to [-90°, 90°]. Besides, an active NPVC method is also proposed for the HVSVM by using one coefficient $k_1$ to adjust two pairs of redundant small vectors. It can remove the dc offset and ac ripples of NPV for full range of $m$ and $\varphi$. The correctness and effectiveness of the proposed HVSVM and active NPVC method are verified by experimental results.

## REFERENCE

[1] A. Nabae, I. Takahashi and H. Akagi, “A new neutral-point-clamped PWM inverter,” *IEEE Trans. Ind. Appl.*, vol. IA-17, no. 5, pp. 518-523, Sept. 1981.

[2] H. Wang, Y. Liu and P. C. Sen, “A neutral point clamped multilevel topology flow graph and space NPC multilevel topology,” in *2015 IEEE Energy Convers. Congr. Expo.*, Montreal, QC, 2015, pp. 3615-3621.

[3] X. Wu, G. Tan, and G. Yao et al., “A hybrid PWM strategy for three-level inverter with unbalanced dc links,” *IEEE J. Emerg. Sel. Top. Power Electron.*, vol. 6, no. 1, pp. 1-15, Mar. 2018.

[4] N. Beniwal, G. Farivar, and H. Lam et al., “Dual-layer pulsewidth modulation technique for average neutral point current control in neutral-point-clamped converters,” *IEEE Trans. Power Electron.*, vol. 37, no. 10, pp. 11762-11773, Oct. 2022.

[5] L. Mihalache, “A hybrid 2/3 level converter with minimum switch count,” in *Conf. Rec. IEEE Ind. Appl. Conf. 41st IAS Annu. Meeting*, 2006, pp. 611-618.

[6] J. Muniz, E. Silva, and R. Nóbrega et al., “An improved pulse-width-modulation for the modified hybrid 2/3-level converter,” *2013 Brazilian Power Electronics Conference*, 2013, pp. 248-253.

[7] P. Najafi, A. Viki, M. Shahparasti, S. Seyedalipour, and E. Pouresmaeil, “A novel space vector modulation Scheme for a 10-Switch converter,” *Energies*, vol. 13, no. 7, pp. 1855-1873, Apr. 2020.

[8] X. Zhu, H. Wang, and X. Deng et al., “Coupled three-phase converter concept and an example: a coupled ten-switch three-phase three-level inverter,” *IEEE Trans. Power Electron.*, vol. 36, no. 6, pp. 6457-6468, Jun. 2021.

[9] X. Deng, H. Wang, and X. Zhu et al., “Common-mode voltage reduction and neutral-point voltage control using space vector modulation for coupled ten-switch three-phase three-level inverter,” *IEEE Trans. Power Electron.*, vol. 37, no. 6, pp. 6397-6411, Jun. 2022.

[10] A. Narendrababu and P. Agarwal, “Virtual vector modulated hybrid 2/3-level Z-source VSI for PV applications,” in *2018 9th IEEE International Symposium on Power Electronics for Distributed Generation Systems (PEDG)*, 2018, pp. 1-7.

[11] X. Deng, X. Zhu, and H. Wang et al., “A novel virtual space vector modulation with optimized neutral-point voltage control capability for ten-switch three-phase three-level inverter,” *IEEE Trans. Ind. Electron.*, vol. 71, no. 2, pp. 1081-1092, Feb. 2024.

[12] X. Deng, H. Wang, and X. Zhu et al., “Fast and accurate neutral-point voltage control under balanced and unbalanced neutral-point voltage for coupled ten-switch three-phase three-level inverter,” *IEEE Trans. Ind. Electron.*, vol. 71, no. 7, pp. 6579-6590, Jul. 2024.